\documentclass{appolb}
\usepackage{epsfig}
\usepackage{cite}
\begin{document}
\pagestyle{plain}
\title{TAU POLARISATION AND ITS CORRELATIONS AS A SIGNAL FOR HIGGS BOSONS 
- UNIVERSAL SPIN INTERFACE FOR TAUOLA PACKAGE
\thanks{Presented at the XXV International School of Theoretical Physics \\
``{\sl Particles and Astrophysics -  Standard Models and Beyond}''  Ustro\'n, Poland,
September 10-16, 2001.} 
~\thanks{Based on work in collaboration with T. Pierzcha\l a, E. Richter W\c as
 and Z. W\c as.}
}
\author{Ma\l gorzata Worek
\address{Institute of  Physics, University of Silesia \\ Uniwersytecka 4,
 40-007 Katowice, Poland \\ e-mail: {\tt Malgorzata.Worek@phys.us.edu.pl}}
}
\maketitle
\begin{abstract}
We show how the $\tau^{+}\tau^{-}$ spin correlations can be used to improve 
the recognition
of the parent boson spin, and hence to identify scalar boson $H^{0}\to \tau^{+}\tau^{-}$ events from the vector boson $Z/\gamma^{*}\to\tau^{+}\tau^{-}$ background in high energy accelerator experiments.

\end{abstract}
%\PACS{14.60.Fg}
\section{Introduction}
The most unsatisfactory feature of the Standard Model is our lack of 
knowledge of the actual mechanism that breaks the electroweak gauge symmetry 
and generates the particle masses. In the Standard Model the breaking occurs 
by a complex Higgs boson doublet. Three components of this doublet become the 
longitudinal polarisation states of the massive vector gauge bosons 
($W^{\pm}$, $Z$), while 
the remaining component manifests itself as a neutral massive scalar Higgs 
particle. On the other hand the minimal supersymetric model contains two 
Higgs boson doublets. Three fields are taken by the vector bosons 
and remaining five become physical particles: a pair of charged 
boson $H^{\pm}$, two scalars
$h^{0}$, $H^{0}$ and one pseudoscalar $A^{0}$. Many more complicated mass 
generation models have been proposed, but their common feature is that the
 couplings of the Higgs bosons to fermions increase with the fermion mass.
Thus for leptons the $H\tau^{+}\tau^{-}$ coupling dominates over the 
$H\mu^{+}\mu^{-}$, $He^{+}e^{-}$ couplings. The experimental observation 
which breaks $e$, $\mu$, $\tau$, universality by favouring $\tau^{+}\tau^{-}$ 
events could be an indication of the presence of a Higgs scalar. Whenever such
a departure from lepton universality is observed a simple helicity 
correlation test can be performed which will clearly indicate that the 
$\tau^{+}\tau^{-}$ pairs have scalar boson origin as opposed to arising 
from vector boson decay. We will show that correlation of 
polarisations of the outgoing $\tau$ leptons are very different 
for $H^{0}\to\tau^{+}\tau^{-}$  decays from that for 
$Z/\gamma^{*}\to\tau^{+}\tau^{-}$ decays. The distinction arises 
because the vector bosons $Z/\gamma^{*}$ decay into either $+~+$ or 
$-~-$ whereas the scalar Higgs bosons $H$ into $+~-$ or $-~+$, where $+,-$ 
denotes the $\tau$ pairs  spin configurations.
\vspace{-0.5 mm}
\section{Universal interface for  TAUOLA package}
We present the algorithm for the interfacing the $\tau$ lepton
 decay package {\tt TAUOLA} - a library of 
Monte Carlo programs 
to simulate decays of $\tau$ leptons \cite{tauola:1990,tauola:1992,tauola:1993}
 with ``any'' production generator to 
include spin effects in the elementary $Z/\gamma^{*} \to \tau^+\tau^-$ 
process \cite{pierzchala:2001}. 
The approximate spin correlation are 
calculated from the 
information stored in the {\tt HEPEVT} common block \cite{PDG:1998} filled by ``any'' $\tau$
production program as described in Ref.\cite{Golonka:2000iu}.  As a demonstration example the
interface is combined  with the {\tt JETSET} generator, however it should work
in the same manner with the {\tt PYTHIA} \cite{Pythia}, 
{\tt HERWIG} or {\tt ISAJET} generators as well. In fact, such an 
interface can be considered as a separate software project, to some degree
 independent both from
the specific problem of $\tau$ production and its decay.\\
The aim of this interface  is {\it not} to replace
the matrix element calculations, but rather to provide a  method 
of calculating/estimating spin effects in cases when spin effects 
would not be taken care of, at all. Such an approach is limited 
for the treatment of longitudinal spin degrees only and to the case 
of  particle production and decay in the ultra-relativistic limit. 
The approximation consists of reconstructing
information of the elementary $2 \to 2$ body 
process $ f\bar{ f} \to (Z/\gamma^{*})\to \tau^+\tau^- $, burried
 inside multi-body production process such as for example 
$ f\bar{ f} \to gZ$, $ f\bar{ f} \to \gamma Z$,   
$ fg \to f Z$, $ f \gamma\to f Z$ etc. The additional 
particles are grouped (summed) into effective quarks and leptons to 
minimise their virtualities. Such an approach is internally consistent 
in the case of photon or gluon emission within the leading log 
approximation. The principle of calculating kinematic variables is simple. 
The 4-momenta of the  $2 \to 2$ body process have to be found.
The 4-momenta of the outcoming $\tau$'s are at present used
 directly~\footnote{This part of the algorithm will be improved in a near
future.}. 
Initial state momenta are constructed from the incoming and outcoming 
momenta of  particles  accompanying production of the 
$Z/\gamma^{*}$ state~\footnote{The $Z/\gamma^{*}$ state  does not need 
to be explicitly coded in the {\tt HEPEVT} common block.}.
Longitudinal polarisation of $\tau$ leptons 
 ${\cal P}_{\tau}$ depends on the spin quantum number
of the $\tau$ mother. It is randomly generated 
as specified in Table~\ref{T:Probability}. 
\begin{table}
\newcommand{\lstrut}{{$\strut\atop\strut$}}
  \caption {\em Probability for the configurations of the longitudinal polarisation of the pair of  $\tau$ leptons from different origins. 
\label{T:Probability}}
\vspace{2mm}
\begin{center}
\begin{tabular}{c c c c} \hline \hline 
& & & \\
Origin & ${\cal P}_{\tau^{+}}$  & $ {\cal P}_{\tau^{-}}$ & Probability \\
& & & \\ 
\hline 
	& & & \\
 Neutral Higgs bosons: $h^{0},H^{0},A^{0}$ &   
 ${\cal P}_{\tau^{+}}=+1$ & ${\cal P}_{\tau^{-}}=-1$  & 0.5 \\
& ${\cal P}_{\tau^{+}}=-1$ & ${\cal P}_{\tau^{-}}=+1$  & 0.5 \\ 
 Charged Higgs boson: $ H^{+}$ or $H^{-}$ &
 ${\cal P}_{\tau^{+}}=+1$ & ${\cal P}_{\tau^{-}}=+1$  & 1.0 \\ 
 Charged vector boson:  $W^{+}$ or $W^{-}$ &
 ${\cal P}_{\tau^{+}}=-1$ & ${\cal P}_{\tau^{-}}=-1$  & 1.0  \\ 
 Neutral vector boson: $Z/\gamma^{*}$ &
 ${\cal P}_{\tau^{+}}=+1$ & ${\cal P}_{\tau^{-}}=+1$  & $P_{Z}$ \\
& ${\cal P}_{\tau^{+}}=-1$ & ${\cal P}_{\tau^{-}}=-1$  & $1-P_{Z}$  \\ 
 Other  &
 ${\cal P}_{\tau^{+}}=+1$ & ${\cal P}_{\tau^{-}}=+1$  & 0.5 \\
& ${\cal P}_{\tau^{+}}=-1$ & ${\cal P}_{\tau^{-}}=-1$  & 0.5 \\
\end{tabular}
\end{center}
\end{table}
The probability $P_{Z}$ used in the generation, 
is calculated directly from the squares 
of the matrix elements of  the Born-level 
$2 \to 2 $ process
$f \bar f  \to \tau^{- }{\tau}^{+ }$:
\begin{equation}
P_{Z}=\frac{\left|{\cal M}\right|^2_{f \bar{f} \to \tau^{-}
\tau^{+ }}\left(+ ,+\right)}
{\left|{\cal M}\right|^2_{f \bar{f} \to \tau^{-} \tau^{+ } }
\left(+ ,+\right)\ +\ 
\left|{\cal M}\right|^2_{f \bar{f} \to \tau^{-} \tau^{+ }  }
\left(- ,-\right)}\; ,
\end{equation}
where $f=e,\mu,u,d,c,s,b$. It can be also expressed
(following conventions of Ref.~\cite{Eberhard:1989ve}), 
with help of the couplings of fermions to the 
$\gamma $ ( and $Z$) bosons. 
\def\Born{{\rm Born}}
\begin{eqnarray}
P_{Z}(s,\theta )&=& {
{d\sigma_\Born \over d\cos\theta }
     \big(s,\cos\theta;1 \big)
\over
{d\sigma_\Born \over d\cos\theta }
     \big(s,\cos\theta;1 \big) +
{d\sigma_\Born \over d\cos\theta }
     \big(s,\cos\theta;-1 \big)
}
\\
{d\sigma_\Born \over d\cos\theta }
     \big(s,\cos\theta;p \big) &=& 
      (1+\cos^2\theta) F_0(s) +2\cos\theta F_1(s) \nonumber \\
&&-p\big[(1+\cos^2\theta) F_2(s) +2\cos\theta F_3(s)\big].
\end{eqnarray}
with the four form-factors
\def\qel{q_f} \def\qta{q_\tau}
\def\vel{v_f} \def\vta{v_\tau}
\def\ael{a_f} \def\ata{a_\tau}
\begin{eqnarray}
F_0(s)&=& {\pi\alpha^2\over 2s}
 \big( \qel^2\qta^2
       + 2\hbox{Re}\chi(s)   \qel\qta\vel\vta
       + \vert\chi(s)\vert^2 (\vel^2+\ael^2)(\vta^2+\ata^2)
  \big),
\nonumber \\
F_1(s)&=& {\pi\alpha^2\over 2s}
 \big( \qquad
         2\hbox{Re}\chi(s)   \qel\qta\ael\ata
       + \vert\chi(s)\vert^2 \; 2\vel\ael \; 2\vta\ata
  \big),
\\
F_2(s)&=& {\pi\alpha^2\over 2s}
 \big( \qquad
         2\hbox{Re}\chi(s)   \qel\qta\vel\ata
       + \vert\chi(s)\vert^2 (\vel^2+\ael^2)\; 2\vta\ata
  \big),
\nonumber \\
F_3(s)&=& {\pi\alpha^2\over 2s}
 \big( \qquad
         2\hbox{Re}\chi(s)   \qel\qta\ael\vta
       + \vert\chi(s)\vert^2 \; 2\vel\ael \; (\vta^2+\ata^2)
  \big),\nonumber 
\end{eqnarray}
and
\begin{equation}
\chi(s)= {s\over s-M_Z^2 +is\Gamma_Z/M_Z } \nonumber  .
\end{equation}
The $q_f$, $v_{f}$, $ a_{f}$, $q_{\tau}$, $v_{\tau}$, $a_{\tau}$ are the
charges and $Z$ coupling constans of the fermions and $\tau$ respectively.

%*******************************************************************
\begin{figure}[h!]
%\centering
\setlength{\unitlength}{0.1mm}
\begin{picture}(1600,800)
%\put( 0,0){\framebox( 1600,800){ }}
\put( 375,750){\makebox(0,0)[b]{\large }}
\put(1225,750){\makebox(0,0)[b]{\large }}
\put(-200, -220){\makebox(0,0)[lb]{\epsfig{file=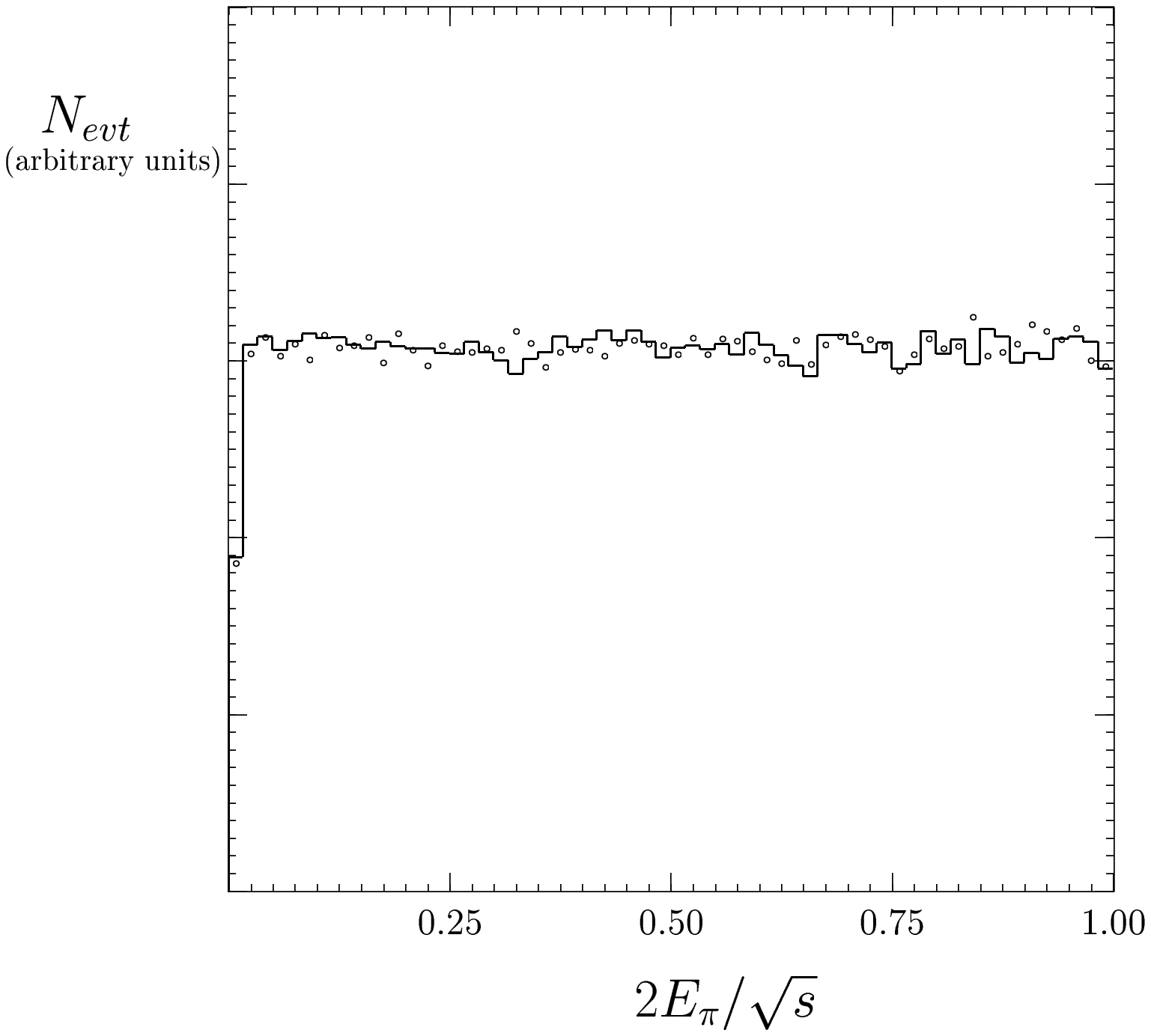,width=90mm,height=120mm}}}
\put(480, -220){\makebox(0,0)[lb]{\epsfig{file=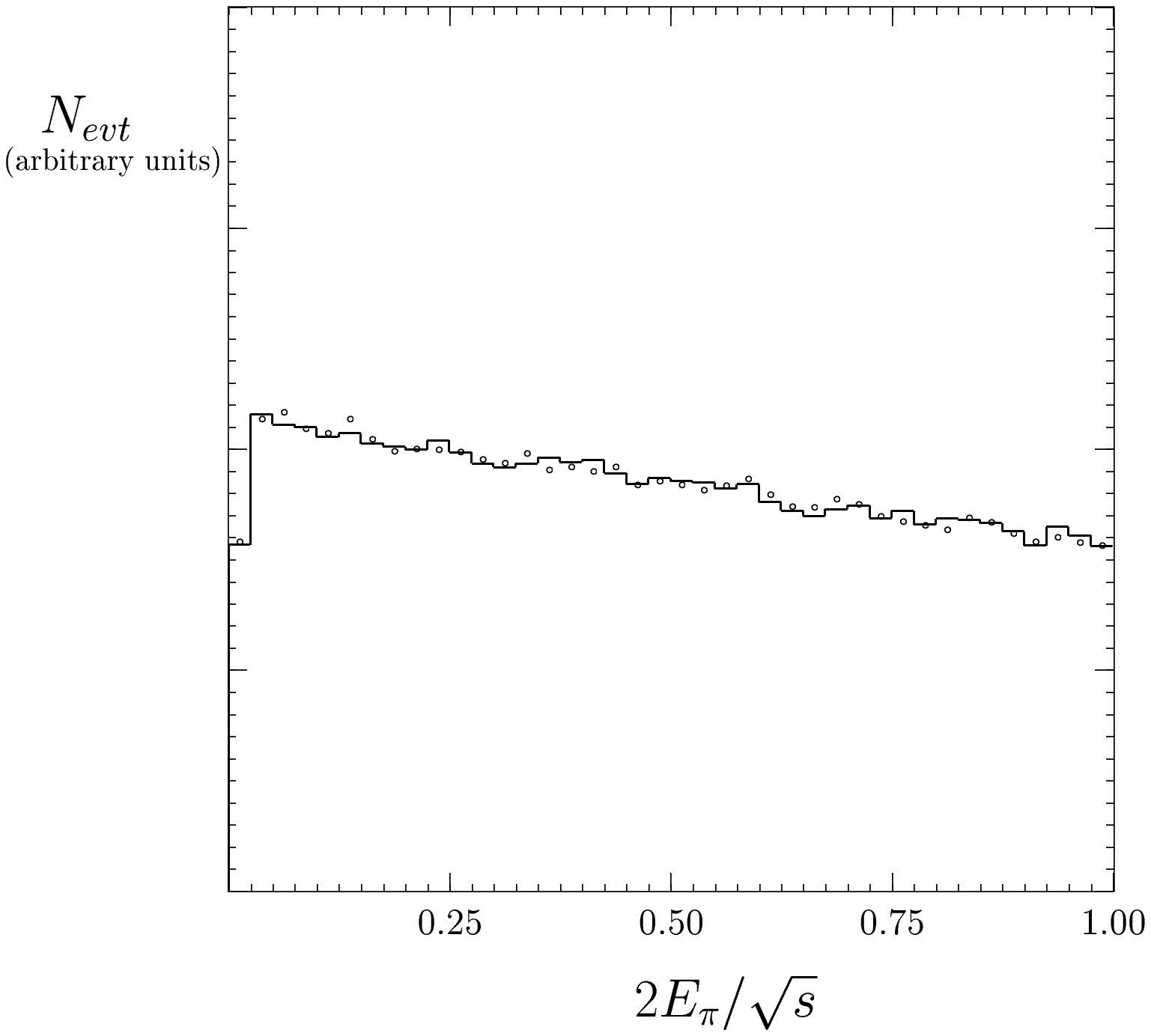,width=90mm,height=120mm}}}

\end{picture}
\caption
{\it Single $\pi$ energy spectrum in  the case of $\tau$ produced from $H$ 
(left-hand  side)  or $Z$ (right-hand  side),
$\sqrt{s} = m_{H}$ or $\sqrt{s} = m_{Z}$ respectively.}
\label{pion1}
\end{figure}
%*************************************************************************
%*******************************************************************
\begin{figure}[!ht!]
%\centering
\setlength{\unitlength}{0.1mm}
\begin{picture}(1600,800)
%\put( 0,0){\framebox( 1600,800){ }}
\put( 375,750){\makebox(0,0)[b]{\large }}
\put(1225,750){\makebox(0,0)[b]{\large }}
\put(-200, -220){\makebox(0,0)[lb]{\epsfig{file=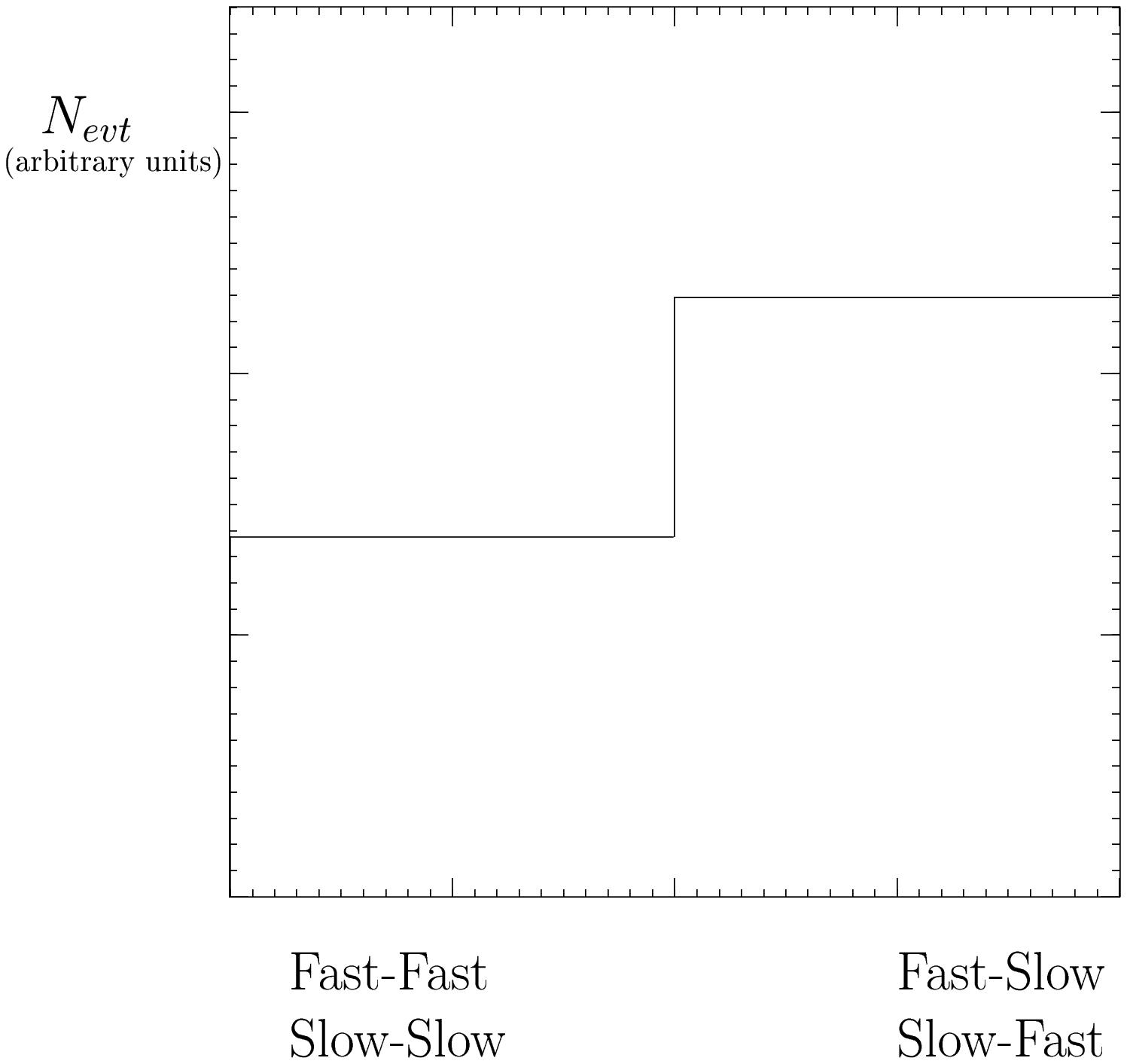,width=90mm,height=120mm}}}
\put(480, -220){\makebox(0,0)[lb]{\epsfig{file=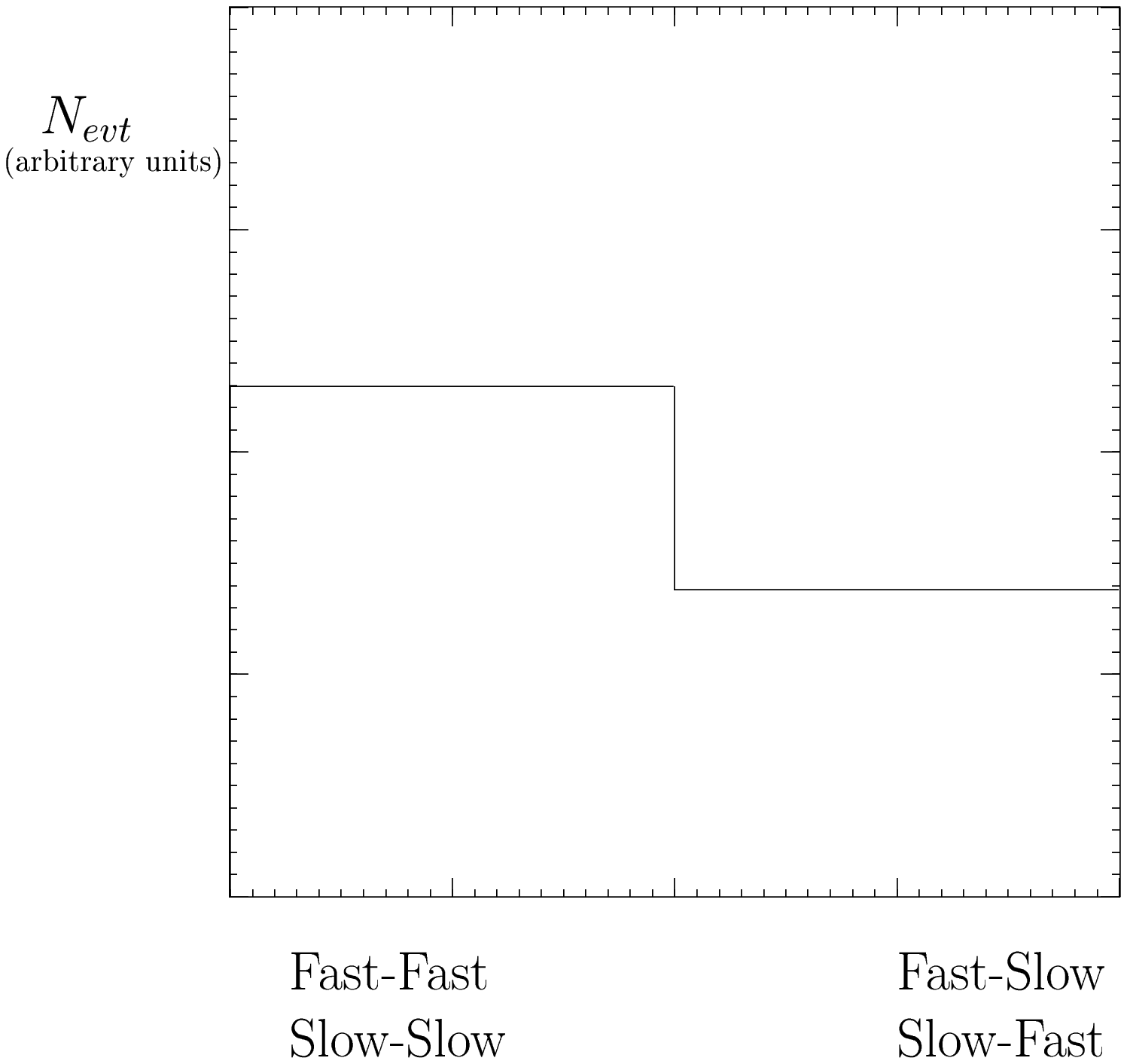,width=90mm,height=120mm}}}

\end{picture}
\caption
{\it $\pi^{+}\pi^{-}$ energy-energy correlations in the case of $\tau$ produced from $H$ 
(left-hand  side)  or $Z$ (right-hand  side),
$\sqrt{s} = m_{H}$ or $\sqrt{s} = m_{Z}$ respectively.}
\label{pion2}
\end{figure}
%************************************************************************* 
%*****************************************************************************
\begin{figure}[!ht]
\setlength{\unitlength}{0.1mm}
\begin{picture}(1600,800)
%\put( 0,0){\framebox( 1600,800){ }}
\put( 375,750){\makebox(0,0)[b]{\large }}
\put(1225,750){\makebox(0,0)[b]{\large }}
\put(-80, -1){\makebox(0,0)[lb]{\epsfig{file=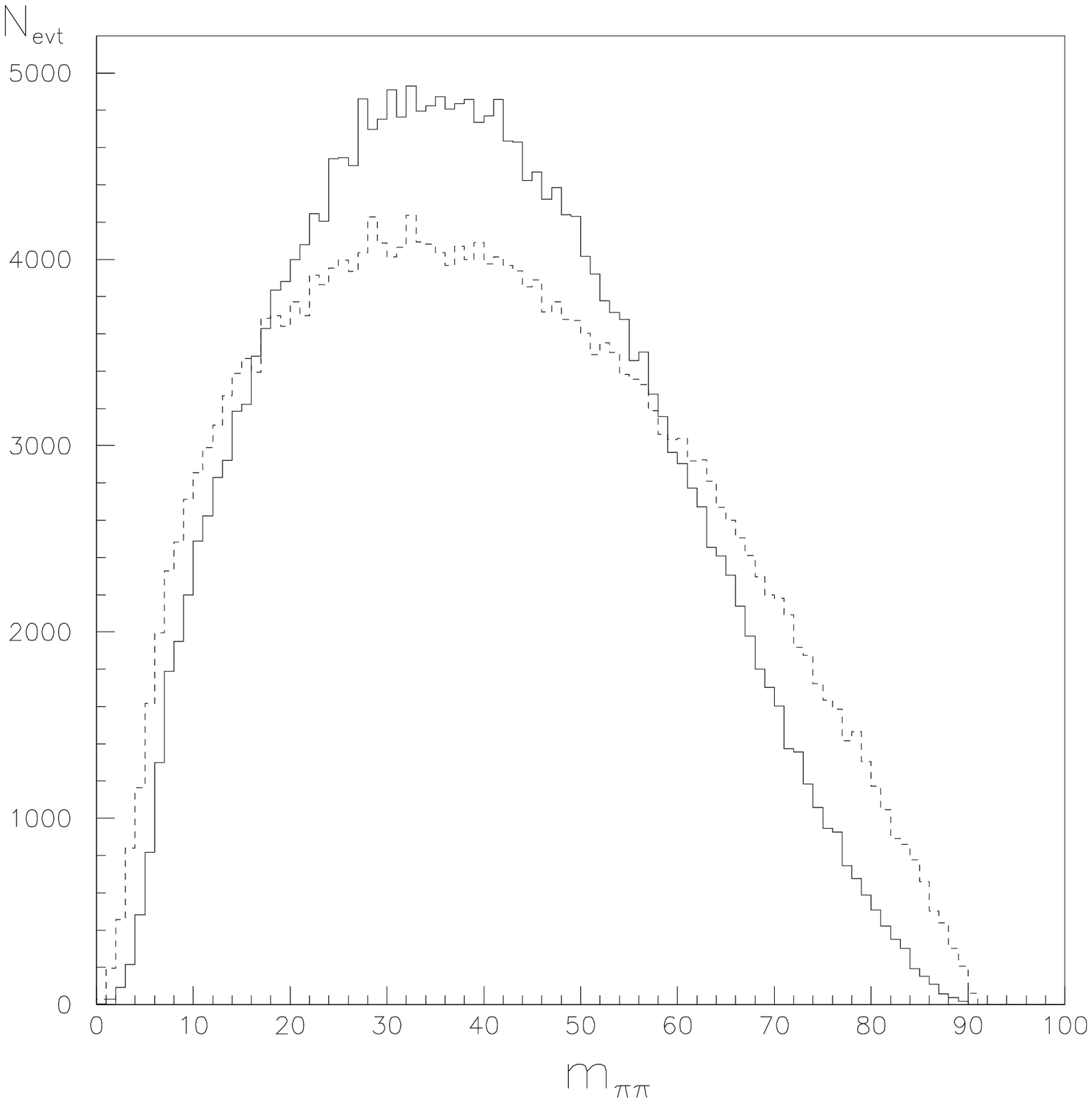,width=75mm,height=75mm}}}
\put(600, -1){\makebox(0,0)[lb]{\epsfig{file=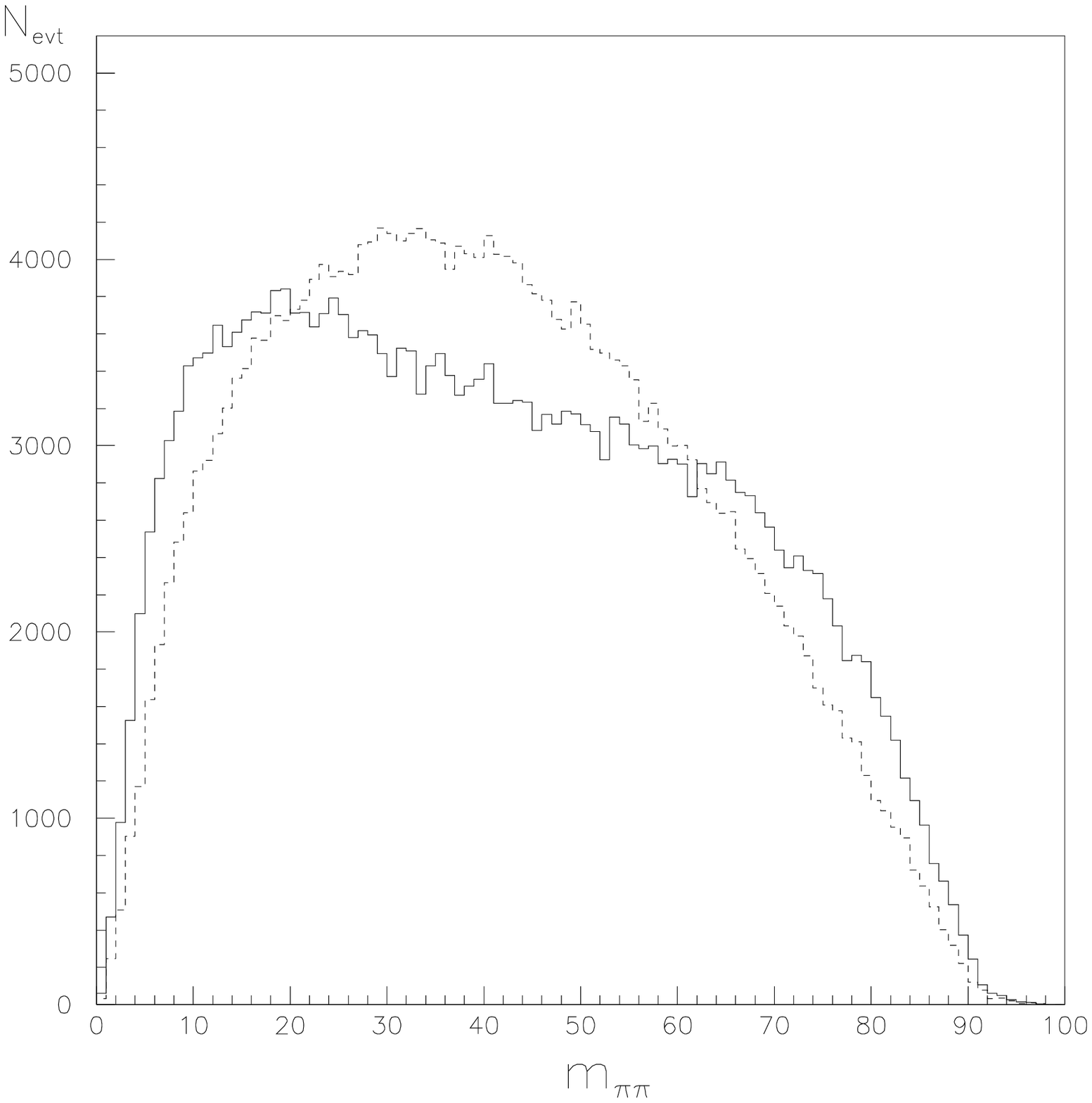,width=75mm,height=75mm}}}
\end{picture}
\caption 
{\it  The $\pi^{+}\pi^{-}$ invariant
 mass distribution. Left-hand side  plot for $H$; right-hand side for $Z/\gamma^*$.  
Continuous line with spin effects included, dotted line with spin 
effects switched off. In the two cases respectively $\sqrt{s}=m_H= m_Z$.   }
\label{invariantmass90}
\end{figure}
%****************************************************************************
%*****************************************************************************
\begin{figure}[!h!]
\setlength{\unitlength}{0.1mm}
\begin{picture}(1600,800)
%\put( 0,0){\framebox( 1600,800){ }}
\put( 375,750){\makebox(0,0)[b]{\large }}
\put(1225,750){\makebox(0,0)[b]{\large }}
\put(-80, -1){\makebox(0,0)[lb]{\epsfig{file=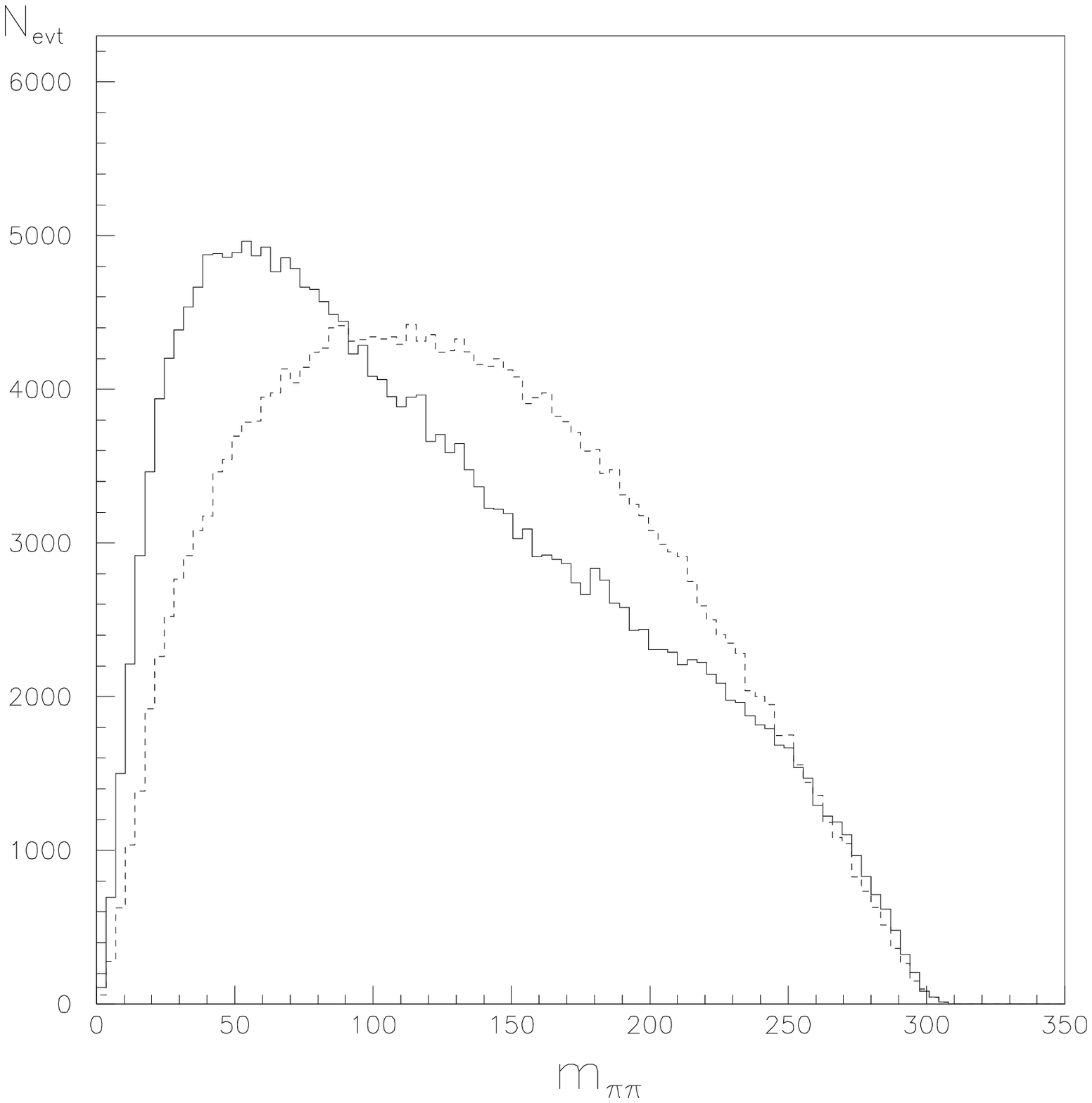,width=75mm,height=75mm}}}
\put(600, -1){\makebox(0,0)[lb]{\epsfig{file=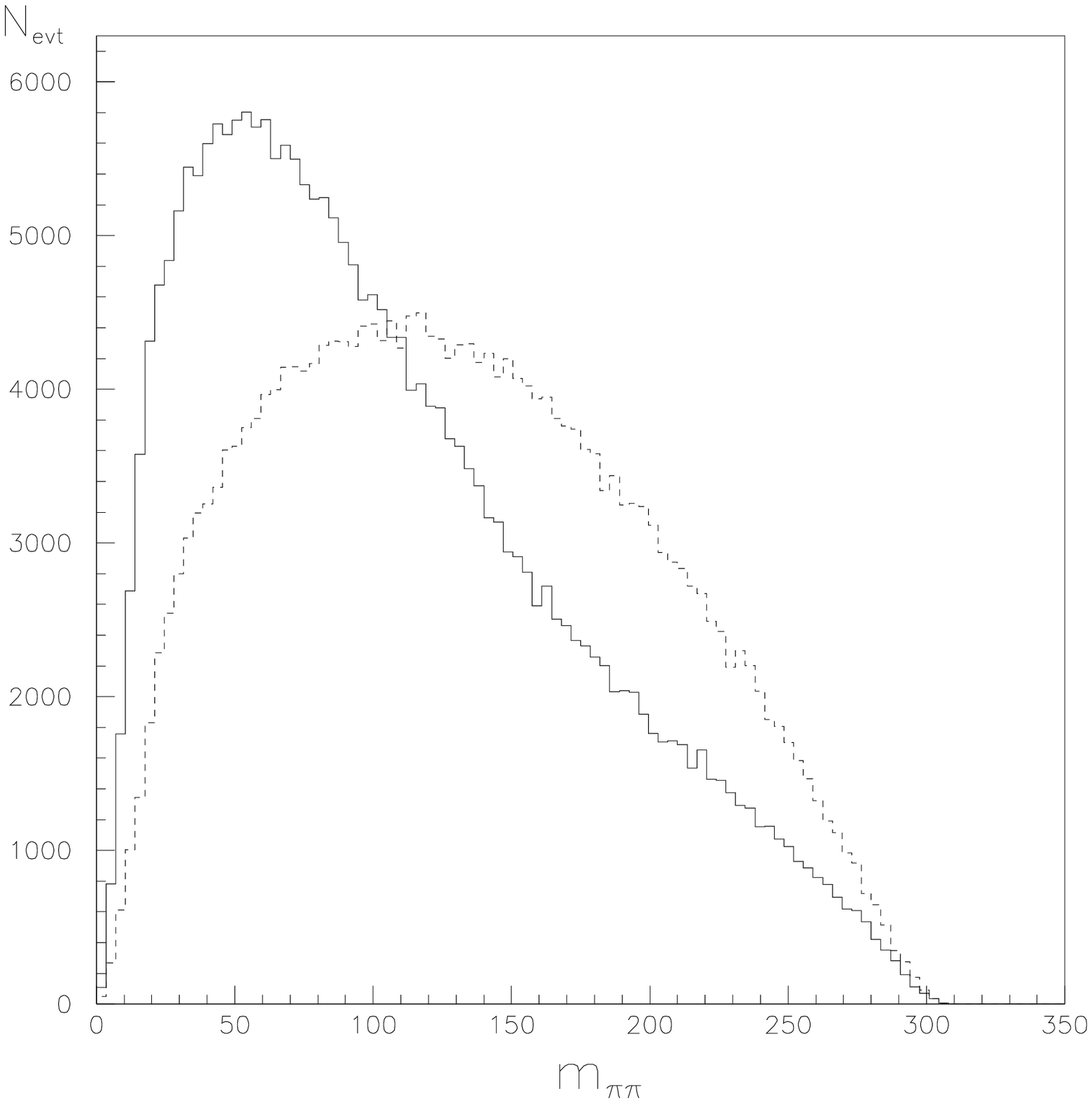,width=75mm,height=75mm}}}
\end{picture}
\caption 
{\it  The $\pi^{+}\pi^{-}$ invariant
 mass distribution for $u \bar u \to Z/\gamma^*$ (left-hand plot) and
$d \bar d \to Z/\gamma^*$ (right-hand plot) produced with cms energy of 300~GeV.  
Continuous line with spin effects included, dotted line with spin 
effects switched off.   }
\label{invariantmass300}
\end{figure}
%****************************************************************************

%*****************************************************************************
\begin{figure}[!ht]
\centering
\setlength{\unitlength}{0.1mm}
\begin{picture}(1600,800)
%\put( 0,0){\framebox( 1600,800){ }}
\put( 375,750){\makebox(0,0)[b]{\large }}
\put(1225,750){\makebox(0,0)[b]{\large }}
\put(-350, -1){\makebox(0,0)[lb]{\epsfig{file=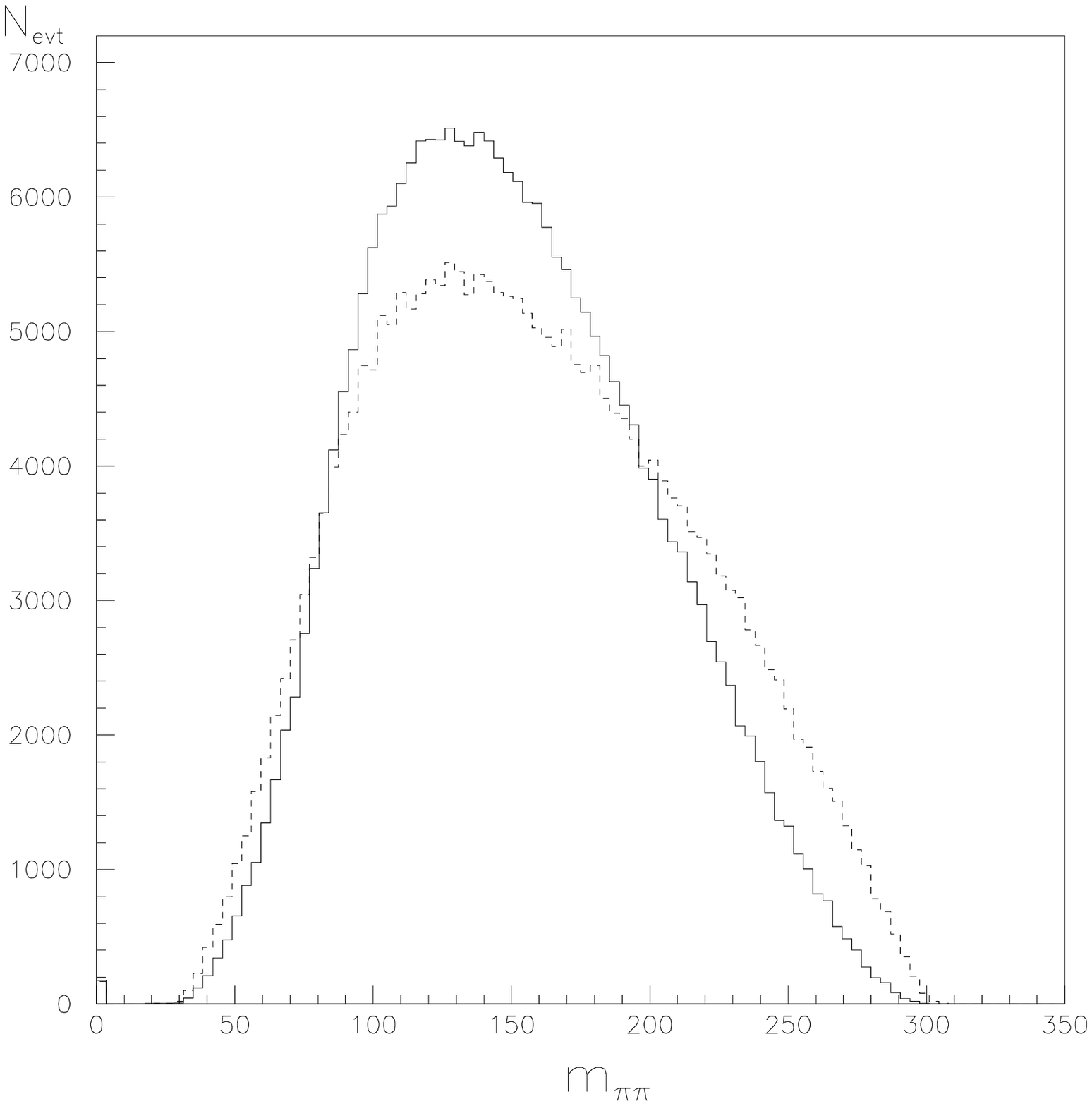,width=75mm,height=75mm}}}
\put(340, -1){\makebox(0,0)[lb]{\epsfig{file=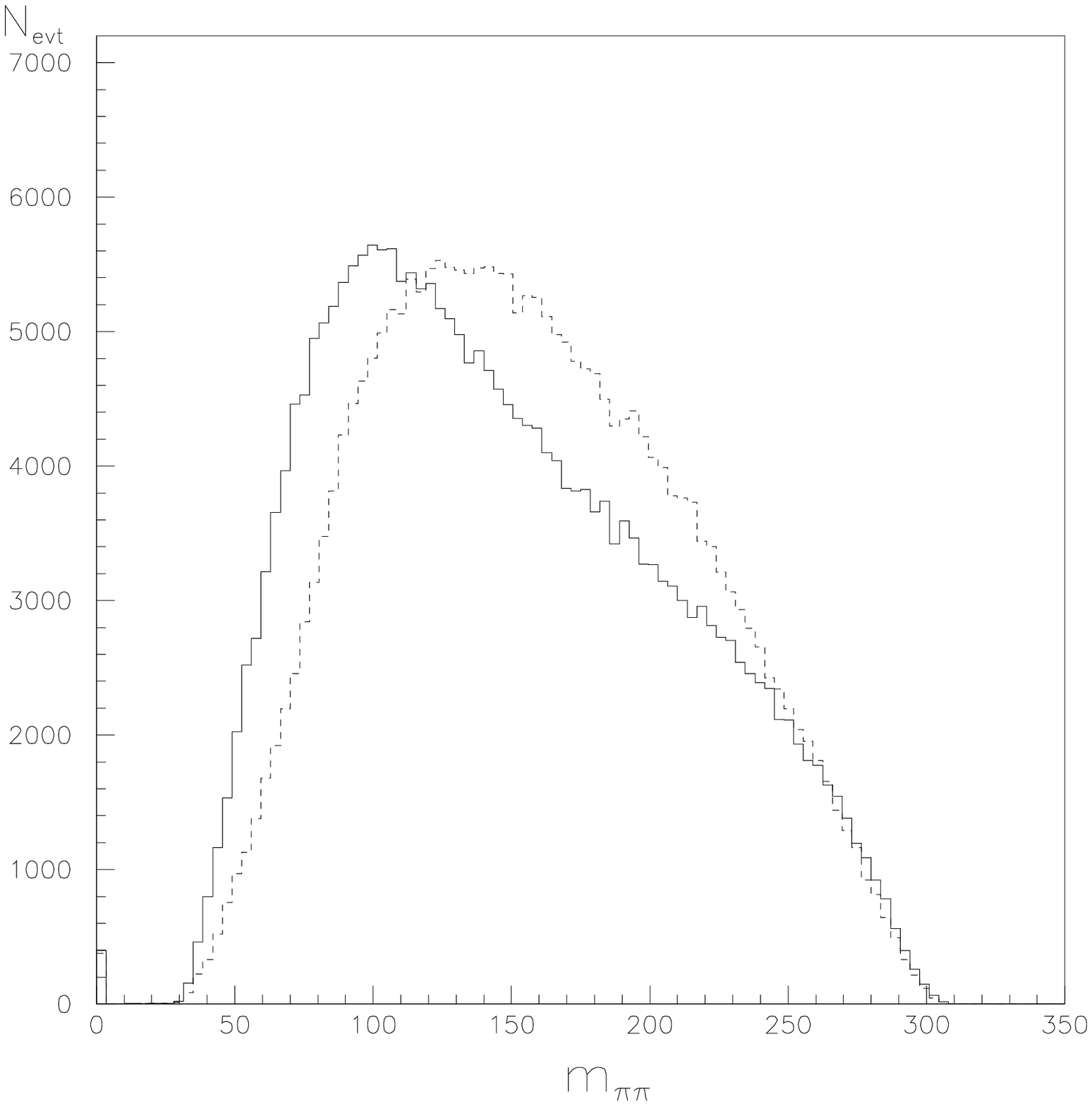,width=75mm,height=75mm}}}
\end{picture}
\caption
{\it The $\pi^{+}\pi^{-}$ invariant
 mass distribution after basic selection. 
On the left  for Higgs boson, on the right for $Z/\gamma^{*}$.
Continuous line with spin effects included, dotted line with spin 
effects switched off. The Higgs boson mass was assumed to be 300~GeV (see text).}
\label{cut5}
\end{figure}
%***************************************************************************
\section{  Spin sensitive observables}
The polarisation of the $\tau$ lepton can be exploited to identify a 
neutral Higgs bosons via the decay $H^{0}\to\tau^{+}\tau^{-}$.
Any experimental observation which breaks $e,\mu,\tau$ universality
- the equality of  $e$, $\mu$ and  $\tau$ couplings to the gauge bosons 
- by favouring $\tau^{+}\tau^{-}$  events could be an indication of 
the presence of a neutral Higgs scalar. The couplings of the ``Higgs'' 
particles to fermions increase with the mass of the fermion, thus the 
$\tau$ couples
 preferentially in comparison with either $\mu$ or $e$.
Here the background is $Z/\gamma^*\to\tau^{+}\tau^{-}$ decays. Whenever such a 
departure from lepton universality is observed there exists a simple 
polarisation correlation test which, if used will help to 
indicate the
 presence of $\tau$ pairs of Higgs boson origin among the background 
$Z/\gamma^*\to\tau^{+}\tau^{-}$ events.  As we can see
from Table~\ref{T:Probability}, the $\tau$ pairs are produced
with the well defined spin configurations:
 $+~+$ or $-~-$  for vector bosons
 $+~-$ or  $-~+$ for neutral Higgs boson. 
Thus a polarisation correlation test can be performed using the energy
distributions of the final decay products which are sensitive to the admixture
of the $H^{0}$ to $Z/\gamma^*$ parentage of the $\tau$ pairs. 
The polarisation correlation can be studied using the various $\tau$ decay
modes.
 Let us concentrate on the 
case of $\tau$ decays to $\pi\nu$, most sensitive to the spin correlations.
The leptonic decay mode \ie  $\tau^{-}\to e^{-}\bar{\nu}_{e}\nu_{\tau}$,
$\tau^{-}\to \mu^{-}\bar{\nu}_{\mu}\nu_{\tau}$  are not
 very sensitive to the polarisation correlation because of the two missing
 neutrinos in this decay. In the case of other hadronic $\tau$ decay, \eg 
$\tau^{-}\to\nu_{\tau}\rho^{-}\to\nu_{\tau}\pi^{-}\pi^{0}$,
 $\tau^{-}\to\nu_{\tau}a_{1}^{-}\to\nu_{\tau}\pi^{-}2\pi^{0}$, 
$\tau^{-}\to\nu_{\tau}K^{*-}\to\nu_{\tau}K^{-}\pi^{0}$ {\it etc.} the
 reconstruction of $\pi^{0}$ is also necessary. 
For the production of the $\tau$ lepton pairs
Monte Carlo program {\tt PYTHIA} was used, and for the decay Monte Carlo 
program {\tt TAUOLA}, and this spin interface \cite{pierzchala:2001}.
It was assured, that the invariant mass of the pair of two incoming quarks
was $\sqrt{s}=m_{Z}= m_{H}$. Energies are
defined in the $\tau^+\tau^-$ pair rest-frame.
With the help of variables $z_\pm= 2E_{\pi^\pm}/\sqrt{s}$, the spin 
effects are visualized.
Fig.~\ref{pion1} shows the  slope 
of $\pi$ energy spectrum (in the case of $Z$)  due to the
$\tau$ polarisation. The slope of the distribution is simply proportional 
to the polarisation. 
\begin{equation}
{d\sigma \over d z_\pm} \sim 1 +{\cal P}_\tau \; 2\; (z_\pm-0.5)
\end{equation}
In the case of the plot on the left-hand side the 
spectrum is flat,
as would be in the case of scalar neutral Higgs boson where
there is no polarisation.
As we can see in the Fig.~\ref{pion2} ($\pi^{+}\pi^{-}$ energy-energy correlations)
 in $Z/\gamma^*\to\tau^{+}\tau^{-}$ decays 
a Fast (a Slow) $\pi^{\pm}$ is most likely to be associated with a 
a Fast (a Slow) $\pi^{\mp}$ whereas the opposite is favoured for  
 $H^{0}\to\tau^{+}\tau^{-}$ decays. 
Therefore, a excess arising from Higgs boson decay can be recognised in the 
$\pi^{+}\pi^{-}$ mode as a Fast $\pi^{\pm}$ with a Slow $\pi^{\mp}$.
The quantity which 
can be measured  experimentally is the invariant mass 
distribution. Fig.~\ref{invariantmass90} shows $\pi^+\pi^-$ invariant mass 
distribution for the  Higgs boson and $Z$ cases.
Continuous line - with spin effects included, dotted line - with spin 
effects switched off. Left-hand side  plot corresponds to the Higgs boson case,
right-hand side to the $Z$.
In the case of Higgs boson, the mass distribution is peaked centrally, whereas in the case of $Z/\gamma^*$ shoulders of the distributions are more profounded. In the $Z/\gamma^*$ case the Fast-Fast and Slow-Slow configurations 
are localised mostly at the shoulders of the  $\pi^+\pi^-$
invariant mass distributions, while for the the Higgs boson case 
the  Fast-Slow configurations are localised in the centre of the 
distributions. If all polarisation effects are switched 
off (dashed lines) the distributions in the two cases are identical.
This observable, {\it i.e.} a
well defined  distribution of  invariant mass built from  
the visible decay products of the $\tau$'s,
can be helpful in separating Higgs boson signal from $Z/\gamma^{*}$ background.
The same distribution have also been studied for the off-peak production of 
$Z/\gamma^*$, {\it i.e.} for the larger cms energies. In these cases the 
average polarisation 
is large and negative, also distinct for the $u \bar u$ and $d \bar d$ 
annihilations. This may open a 
way for measuring the flavour of the quarks leading to $\tau$ pair production. 
As illustrated in Fig.~\ref{invariantmass300}, the effect on the 
$\pi^{+}\pi^{-}$ invariant mass distribution is noticeable.
The shape of the distribution might give the insight to the structure 
functions of  colliding protons.
\section{ Case of the Higgs boson signatures at LHC}
The $\tau$ leptons are considered as a very promising signature for the 
searches of the Higgs bosons in the Minimal Supersymmetric Standard Model
 (MSSM) at LHC collider \cite{ATLAS-TDR, CMS-TP}. The neutral Higgs bosons 
H and A decay into 
 $\tau^+ \tau^- $ pair, are enhanced
for the large values of $\tan \beta$
($\tan \beta$ denotes the ratio of the vacuum expectation values of the
 Higgs doublets in the MSSM model),
with the branching ratio of about 10\% for most of the range of the 
interesting Higgs boson mass values ( 150-1000 GeV ).
The irreducible background to this process is a  $Z/\gamma^* \to \tau \tau$ 
decay and the reducible backgrounds is the QCD jets.
In this study the $\tau$ identification is based on the presence of a 
single hard isolated charged hadron in the jet using tracker information.
Two hard tracks from $\tau^{-}$ and $\tau^{+}$ in the signal events have
 an opposite sign while no strong charged correlation is expected for the 
QCD jets.
Fig.~\ref{cut5} shows the effect of  $\pi \pi$ 
invariant mass distribution. Events were generated with the 
Monte Carlo {\tt PYTHIA 5.7}  \cite{Sjostrand:2000wi},
the Higgs boson mass of the 300 GeV and the width below 1 GeV
(as for $\tan \beta~\sim~10$) was assumed. 
For the  $Z/\gamma^*$,
the cms energy of the produced $\tau \tau$ pair was taken in the 
range 300~$\pm$~10~GeV. A simple selection was applied. 
The minimal transverse momenta of the  $\pi's$ were required to be 
above 15~GeV and the pseudorapidity $|\eta|~<~2.5$ \cite{ATLAS-TDR}. 
In Fig.~\ref{cut5} we can see  the visible effect of
spin-correlations. 
Similarily as in case of Fig.~\ref{invariantmass90}, the distribution
for the $Z/\gamma^{*}$ has more profounded shoulders than for the Higgs 
boson, due to spin correlations.
\section{ Summary}
We have discussed the algotithm for interfacing the $\tau$ lepton decay package {\tt TAUOLA} with ``any'' production generator to include effects due to spin in the elementary $Z/\gamma^{*}\to\tau^{+}\tau^{-}$ process 
\cite{pierzchala:2001}.
The invariant mass distributions presented here, sensitive to the
$\tau^+\tau^-$ spin correlations,
can possibly be used for the MSSM Higgs boson searches at LHC
to enhance sensitivity of the signal or to verify the hypothesis
of the spin zero nature of the Higgs boson.
This code is publicly availale from 
the address~\cite{TauolaInterface}.\\ \\

It is a pleasure to thank Tomasz Pierzcha\l a, El\.zbieta Richter-W\c as and 
Zbigniew W\c as, with whom the work reported here was performed.  
I wish to thank Marek Biesiada for giving me the opportunity to give this talk at the XXV   International School of Theoretical Physics in Ustro\'n. 
This work is partly supported by the Polish State Committee grants KBN 
5P03B10121, 2P03B04919.


\begin{thebibliography}{99}
\bibitem{tauola:1990} S. Jadach, J.H. K\"uhn, Z. W\c{a}s, {\it Comput. Phys. Commun.} 
{\bf 64}, 275 (1990).
\bibitem{tauola:1992} M. Je\.zabek, Z. W\c{a}s, S. Jadach, J.H. K\"uhn 
{\it Comput. Phys. Commun.} 
{\bf 70}, 69 (1992). 
\bibitem{tauola:1993} R. Decker, S. Jadach, J.H. K\"uhn, Z. W\c{a}s,
{\it Comput. Phys. Commun.} 
{\bf 76}, 361 (1993). 
\bibitem{pierzchala:2001}
T. Pierzcha\l{}a, E. Richter-W\c{a}s, Z. W\c{a}s and M. Worek, 
{\it Acta Phys. Pol.} {\bf B32}, 1277 (2001).
\bibitem{PDG:1998} Particle Date Group, C. Caso {\it et al.}, 
{\it Eur. Phys. J.} {\bf C3},1 (1998).
\bibitem{Golonka:2000iu} P. Golonka, E. Richter-W\c{a}s, Z. W\c{a}s, 
{\tt hep-ph/0009302}. 
\bibitem{Pythia} T. Sj\"ostrand, {\it Comput. Phys. Commun.} {\bf 82}, 74 
(1994).
\bibitem{Eberhard:1989ve}P.H. Eberhard {\it et al.}, Proceedings of the
 Workshop on
 $Z$ Physics at LEP,\\ edited by G. Altarelli, R. Kleiss and V. Verzegnassi, 
CERN-89-08 v. 1-3, \\Switzerland, Geneva (1989).
\bibitem{ATLAS-TDR} ALEPH Collaboration, CERN-LHCC/99-15.
\bibitem{CMS-TP} CLEO Collaboration, CERN-LHCC/94-44.
\bibitem{Sjostrand:2000wi} T. Sj\"ostrand {\it et al.}, {\tt hep-ph/0010017}.


\bibitem{TauolaInterface} {\tt www.home.cern.ch/$^{\sim}$wasm}.
\end{thebibliography}
\end{document}